\documentclass[10pt,aps,prd,twocolumn,amssymb,superscriptaddress,nofootinbib,floatfix,longbibliography]{revtex4-1}

\usepackage{graphicx}
\usepackage{amsfonts}
\usepackage{amsmath}
\usepackage{stmaryrd}
\usepackage{hyperref}
\usepackage{units}
\usepackage[latin1]{inputenc}      
\usepackage{color}
\usepackage{dcolumn}
\usepackage{bm}
\usepackage{cleveref}
\usepackage{mathtools}
\usepackage[normalem]{ulem}

\newcommand{\om}{\omega}

\newcommand{\p}{\partial}
\newcommand{\be}{\begin{equation}}
\newcommand{\ee}{\end{equation}}
\newcommand{\bsub}{\begin{subequations}}
\newcommand{\esub}{\end{subequations}}
\newcommand{\bea}{\begin{eqnarray}}
\newcommand{\eea}{\end{eqnarray}}
\newcommand{\bi} {\begin{itemize}}
\newcommand{\ei} {\end{itemize}}
\newcommand{\bmat} {\begin{pmatrix}}
\newcommand{\emat} {\end{pmatrix}}

\renewcommand{\vec}[1]{\mathbf{#1}}

\makeatletter
\let\cat@comma@active\@empty
\makeatother

\begin{document}


\title{Quasinormal mode oscillations in an analogue black hole experiment}

\author{Theo Torres}
\affiliation{School of Mathematical Sciences, University of Nottingham, University Park, Nottingham, NG7
2RD, UK}

\author{Sam Patrick}
\affiliation{School of Mathematical Sciences, University of Nottingham, University Park, Nottingham, NG7
2RD, UK}

\author{Maur\'icio Richartz}
\affiliation{Centro de Matem\'atica, Computa\c c\~ao e Cogni\c c\~ao,
Universidade Federal do ABC (UFABC), 09210-170 Santo Andr\'e, S\~ao Paulo, Brazil}

\author{Silke Weinfurtner}
\affiliation{School of Mathematical Sciences, University of Nottingham, University Park, Nottingham, NG7
2RD, UK}
\affiliation{School of Physics and Astronomy, University of Nottingham, Nottingham, NG7 2RD, UK.}
\affiliation{Centre for the Mathematics and Theoretical Physics of Quantum Non-Equilibrium Systems, University of Nottingham, Nottingham, NG7 2RD, UK.}


\begin{abstract}
\noindent  The late stages of the relaxation process of a black hole are expected to depend only on its mass and angular momentum, and not on the details of its formation process. 
Inspired by recent analogue gravity experiments, which demonstrate that certain black hole processes take place in gravitational and hydrodynamical systems alike, we conduct an experiment to search for quasinormal mode oscillations of the free surface of a hydrodynamical vortex flow. Our results demonstrate the occurrence and hint at the ubiquity of quasinormal ringing in non-equilibrium analogue black hole experiments.

\end{abstract}

\maketitle

\noindent {\bf Introduction and motivation.}
Analogue gravity provides a testing ground for many ideas in general relativity, both theoretically and experimentally~\cite{Barcelo05,Barcelo:2018ynq}. Over the past decade, much work has been directed to probing gravitational phenomena such as Hawking radiation~\cite{Weinfurtner10,Steinhauer15,Euve16,deNova:2018rld} and superradiance~\cite{Torres16} in analogue systems.
In this letter we investigate the relaxation of perturbed systems towards equilibrium via the emission of quasinormal modes, the frequencies of energy dissipation. For the first time, we measure the frequency spectrum of a perturbed hydrodynamical vortex which possesses many features of a rotating black hole. We demonstrate that quasinormal ringing is an integral part of a hydrodynamical non-equilibrium process. The observed spectrum is in excellent agreement with the theoretically predicted frequencies associated with the stationary orbits\footnote{Such orbits are referred to as light-rings in the context of black holes.} of surface waves surrounding an out-of-equilibrium vortex flow. 

The analogy, built on the works of Unruh~\cite{Unruh} and Visser~\cite{Visser93}, relies on the fact that perturbations of certain condensed matter systems, e.g.~shallow water waves propagating on the free surface of an irrotational vortex flow~\cite{Schutzhold}, are mathematically equivalent to scalar waves propagating around a rotating black hole.
Such a vortex flow, commonly called a draining bathtub (DBT) flow, is uniquely described by the velocity field \mbox{$\vec{v}(t,r,\theta) = \vec{v}(r) = -\frac{D}{r} 
\boldsymbol{\hat{r}} + \frac{C}{r} \boldsymbol{\hat{\theta}}$}. The two parameters, denoted $C$ (for circulation) and $D$ (for drain), are analogous to the angular momentum and mass of a rotating black hole. In fact, the region where the flow is sufficiently fast to trap any wave is analogous to the event horizon of a black hole. Similarly, the region where the flow is sufficiently fast that waves are dragged along the flow is analogous to the ergosphere of a rotating black hole.

Once perturbed, a vortex flow will relax through the emission of surface waves that propagate on the air-water interface. Such waves correspond to small deformations, $\delta h(t,\vec{x})$, of the unperturbed surface elevation, where $t$ denotes time and $\vec x$ represents the spatial coordinates on the two-dimensional free surface. 
For an axisymmetric free surface vortex, it is convenient to adopt polar coordinates $\vec x = (r,\theta)$ and decompose the perturbations in terms of its azimuthal components, 
\be \label{eq:ringdown}
\delta h (t,r,\theta) = \mathrm{Re} \left[  \sum_{m \in \mathbb{Z}} \delta h_m(t,r) e^{im\theta} \right] ,
\ee
where the azimuthal number $m \in \mathbb{Z}$ indicates an $m$-fold symmetry with respect to the polar angle $\theta$.

Towards the end of the relaxation process, each azimuthal component is well-approximated (in an open system) by a superposition of time-decaying modes,
called \textit{quasinormal modes} (QNMs)~\cite{Kokkotas99,Berti,Konoplya11}.
Each QNM oscillates at the characteristic frequency $f_ {m n}$ and has an amplitude that decays exponentially in time with a characteristic timescale of $1/\Gamma_{m n}$. 
The overtone number $n \in \mathbb{N}$ classifies  the QNMs according to their decay times.
The set of complex frequencies $\omega_{\rm QNM}(m,n) = 2 \pi f_{mn} + i\Gamma_{mn}$ is called the QNM spectrum. For the DBT flow, the QNMs have been extensively studied and their spectrum was calculated using various methods~\cite{Berti:2004ju,Dolan3}. The QNM spectra of more realistic vortex flows, either due to the presence of vorticity~\cite{flowmaster} or dispersion~\cite{TheoLR}, have also been investigated.

One of the techniques available to estimate the QNM spectrum of a black hole is based on the properties of \textit{light-rings} (LRs)~\cite{Goebel72,Cardoso}. 
The LRs of a black hole are the orbits (i.e.~the equilibrium points in the radial direction) of massless particles. 
The relation between QNMs and LR modes comes from the fact that in many (but not all~\cite{khanna}) spacetimes QNMs can be seen as waves travelling on the unstable orbits and slowly leaking out~\cite{Berti}. 
While QNMs strongly rely on the openness of the system, the LR modes, being independent of the boundary conditions, do not. In particular, 
the presence of a non-open boundary condition, either at infinity or at the horizon, will modify the late-time behaviour of the relaxation process.
More precisely, the decay times of characteristic waves will be altered by reflections from the  boundaries. 

Additionally, their non-oscillatory behaviour will be further modified by damping in the system and by recurring  perturbations.
The oscillatory part of the LR spectrum therefore provides a more robust quantity to characterise the fluid flow in finite size experiments.
By measuring the characteristic frequencies associated with the relaxation of a vortex fluid flow, and by verifying their agreement with theoretical predictions, our experiment provides the first detection of quasinormal oscillations in an analogue black hole experiment. 

\vspace*{0.2cm}

\noindent {\bf \emph Theory.} It is known that the QNM spectrum can be approximated using the properties of the LRs via~\cite{Cardoso}:
\begin{equation}
\omega_{\rm{QNM}}(m,n) \approx \omega_{\star}(m) - i\Lambda(m) \left( n + \frac{1}{2} \right),
\end{equation}
where $\omega_\star(m) = 2 \pi f_\star(m)$ is the angular frequency of an $m$-mode orbiting on the LR and $\Lambda(m)$ is the \textit{Lyapunov exponent} of the orbit for this specific $m$-mode. 
In 2D systems where dispersion is absent, the evolution of perturbations can be described by means of a potential function $V(\omega,m,r)$ and the properties of the LRs can be found in the high-$m$ limit of such potential (see for e.g.~ \cite{Dolan2,Mauricio14,flowmaster}).

 When dispersion is present, however, one does not have access to a potential description. In order to evaluate the LR spectrum in such a case, we look at the trajectories of rays. Such trajectories are found by solving Hamilton's equations with the Hamiltonian~\cite{nazarenko,TheoLR}
\begin{equation}
\mathcal{H}(\vec{x},\vec{k},t,\omega) = - \frac{1}{2} \left(\omega - \vec{v}.\vec{k} \right)^2 + \frac{1}{2} F(k),
\end{equation}
where
\begin{equation}
 F(k) = \sqrt{\left(gk + \frac{\sigma}{\rho} k^3\right)\tanh(hk)},  
\end{equation}
$k = |\vec k |$, $g = 9.81$ $\mathrm{m}/\mathrm{s}^2$ is the gravitational acceleration on Earth, $\sigma=0.0728$ $\mathrm{N}/\mathrm{m}$ is the surface tension of water, $\rho=997$ $\mathrm{kg}/\mathrm{m}^3$ is the density of water, $h$ is the water depth, and $\vec{v}$ is the velocity of the background flow. 
This amounts to a WKB approximation with an associated wave vector given, in polar coordinates, by $\vec{k} = (\frac{m}{r}, k_r)$. 

The LR\footnote{Even though these are surface water waves and not electromagnetic waves, we shall still refer to them as LR modes due to the fluid-gravity analogy.} oscillation frequency, $f_\star$, is calculated by looking for the critical points of the Hamiltonian. More precisely, for a given azimuthal number $m$ and for fixed flow parameters, we solve Hamilton's equations, i.e.~
\begin{equation} \label{LRcond}
\mathcal{H} = 0, \quad \frac{\p \mathcal{H}}{\p r} = 0 \quad \text{ and } \quad \frac{\p \mathcal{H}}{\p k_r} = 0,
\end{equation}
numerically in order to determine the location of the LR, its wavenumber $k_r$ and, most importantly for us, its oscillation frequency $f_\star(m)$.

We note that there is a clear physical interpretation of the set of equations above. First, the condition $\mathcal{H} = 0$ means that surface waves propagating on a background flow satisfy the dispersion relation
\begin{equation} \label{dispersion_relation}
\om = \om_d(\vec{k}) = \pm F(k) + \vec{v}.\vec{k}.
\end{equation}
Second, $\frac{\p \mathcal{H}}{\p k_r} = 0$ is equivalent to $\p \omega_d/\p k_r =0$, which defines two curves, $\omega^{\pm}_{\rm{min}} = 2\pi f^{\pm}_{\rm{min}}(m,r)$ (one for each branch of the dispersion relation - in this work, without loss of generality, we work with the $+$ branch). They represent the minimum frequency required by a specific $m$-mode to be able to propagate at a radius $r$. These curves separate the $(f,r)$-plane in two regions. On the one hand, modes with a frequency $f$ above the minimum energy curve, i.e.~$f > f^{+}_{\rm{min}}$, have a real-valued $k_r$, and are therefore able to propagate. On the other hand, modes with a frequency $f$ below the minimal energy threshold, i.e.~$f < f^{+}_{\rm{min}}$, having imaginary $k_r$ values, correspond to evanescent modes. Finally, $\frac{\p \mathcal{H}}{\p r} = 0$ is equivalent to $\p \omega_d/\p r =0$, implying that LR modes correspond to the lowest energy modes that are able to propagate across the entire fluid. By sitting at the top of the minimum energy curve, the LR modes possess a real valued $k_r$ everywhere, allowing them to transfer energy across the entire system.

\vspace*{0.2cm}

\noindent {\bf \emph Experimental setup.} We set up a vortex flow out of equilibrium to observe its quasinormal oscillations. Our experiment was conducted in a $3~\rm{m}$ long and $1.5~\rm{m}$ wide rectangular tank with a $2~\rm{cm}$-radius sink hole at the centre. Water is pumped continuously from one corner at a flow rate of $15\pm 1~\ell/\rm{min}$. 
The sink-hole is covered until the water raises to a height of $10.00 \pm 0.05~\rm{cm}$. Water is then allowed to drain, leading to the formation of a vortex. We call such a restless fluid flow an \textit{Unruh vortex}\footnote{This name comes from the German word ``Unruhe" which means restlessness and was chosen in acknowledgment of W. G. Unruh, the founder of analogue gravity.}.  
We recorded the perturbations of the free surface when the flow was in a quasi-stationary state at a water depth of $5.55 \pm 0.05~\rm{cm}$. 
The resulting Unruh vortex is axisymmetric to a good approximation and its surface flow is well described by the DBT model in the region of observation, i.e.~in the circular annulus with inner radius $7.4~\mathrm{cm}$ and outer radius $25~\mathrm{cm} $ around the drain hole. By employing a standard \textit{Particle Imaging Velocimetry} (PIV) technique, we estimate the circulation parameter to be $C \approx 151 ~\rm{cm}^2/\rm{s}$ and the drain parameter to be negligible, i.e.~$D \approx 0 ~\rm{cm}^2/\rm{s}$ (see the appendix for details on flow characterisation). The water elevation was recorded using the Fast-Chequerboard Demodulation method~\cite{Sanders} and the entire procedure was repeated 25 times. 

We note that, even though the drain parameter (corresponding to the contribution of the surface flow to the drain rate) is negligible where the DBT model is used, inside the vortex core, where vorticity and other effects become relevant, the drainage from the surface of the fluid is important. In fact, perturbations deep inside the vortex core cannot propagate out of it. In this sense, the vortex flow is an analogue black hole and the vortex core mimics the event horizon of a black hole.

\begin{figure}[!] 
\includegraphics[scale=0.95]{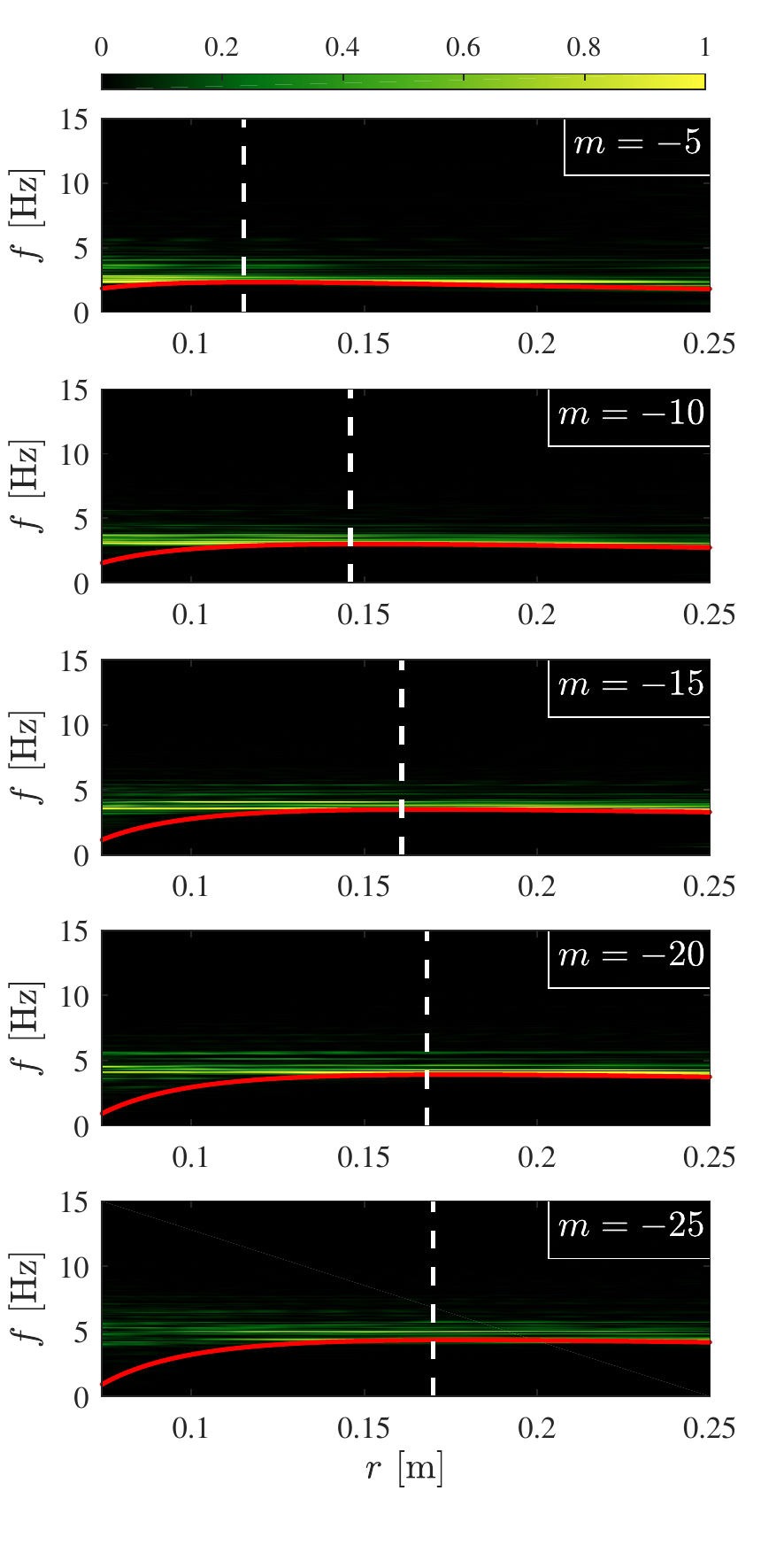} 
\caption{{\bf Normalised power spectral densities}.
The power spectral density is compared with the minimum energy curve $f_{\rm{min}}^{+}(m,r)$, plotted in red, for various $m$. The maxima of $f_{\rm{min}}^{+}(m,r)$ indicate the location of the light-rings, which are shown in dashed white lines. Note that the spectral density peaks and the minimum energy line are distinguishable for small radii.}
\label{psd}
\end{figure}

\vspace*{0.2cm}
\noindent      {\bf \emph Results.}   The axisymmetry of our setup allows us to perform an azimuthal decomposition, as in~\eqref{eq:ringdown}, to study quasinormal oscillations. More precisely, we select specific azimuthal modes by performing a polar Fourier transform and we extract the associated radial profiles $\delta h_m(t,r)$. Azimuthal modes with $m>0$ are co-rotating with the flow while modes with $m<0$ are counter-rotating with the flow. In our experiment, we focus on the counter-rotating modes since the co-rotating modes are associated with LRs located deep inside the vortex core, where the DBT model breaks down.

\begin{figure}[!htp] 
\includegraphics[trim = 0.5cm 0 0 0, scale =0.8]{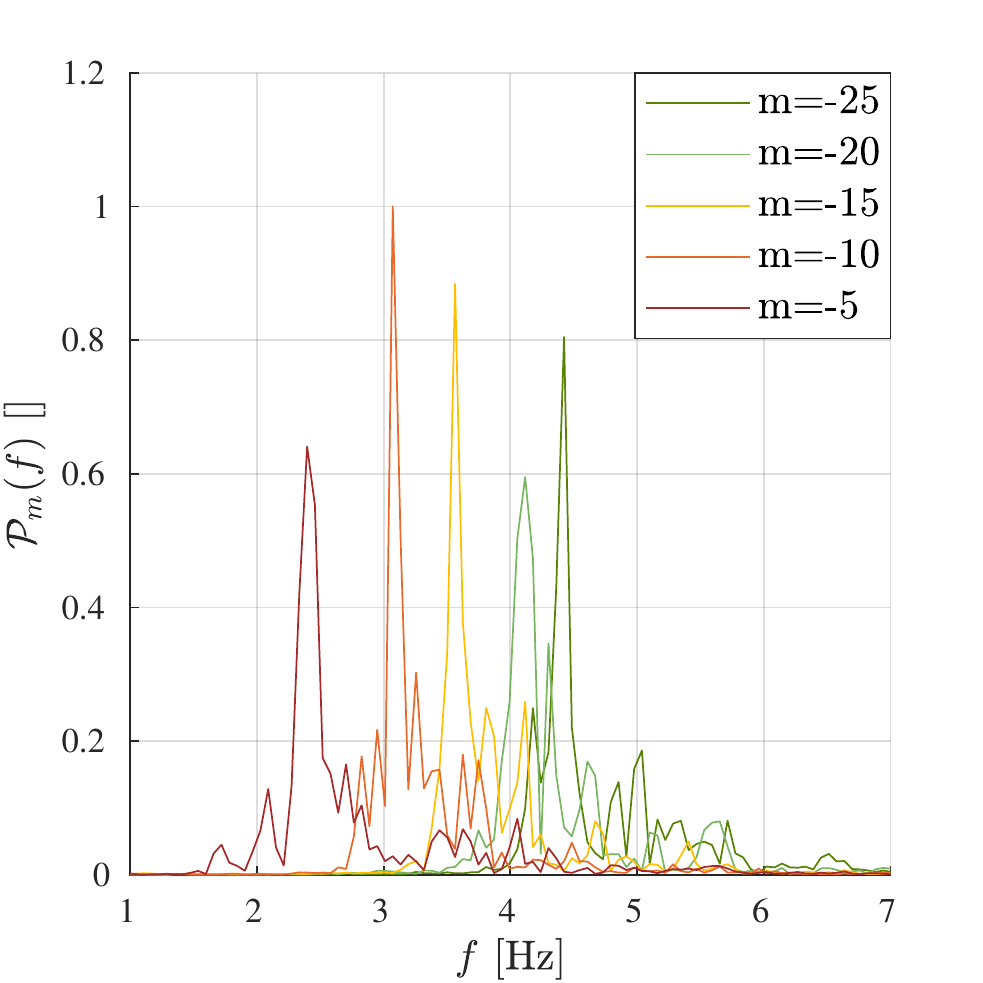} 
\caption{{\bf Typical radius-averaged power spectral densities.} The location of peaks, denoted by $f_{\mathrm{peak}}(m)$, correspond to quasinormal oscillation frequencies. These values of $f_{\mathrm{peak}}(m)$ extracted from the data are plotted in Fig.~\ref{spec}.}
\label{peaks}
\end{figure}

By calculating the time Fourier transform of $\delta h_m(t,r)$, we estimate the \textit{Power Spectral Density} (PSD) of each $m$-mode for 
$r \in [7.4~\rm{cm},25~\rm{cm}]$. 
In Fig.~\ref{psd} we present the PSDs of a single experiment for a range of modes. The minimum energy curves, $f^{+}_{\rm{min}}$, are plotted in red as a function of $r$ for various values of $m$. Note that, unlike the dispersion relation, the PSDs are approximately constant over the window of observation. Taking this into account, we average over the radius in order to obtain the r-independent frequency content. Various averaged PSDs, denoted by $\mathcal{P}_m(f)$, are presented in Fig.~\ref{peaks}. For each $m$ value the spectral density is peaked around a single frequency, which allows us to extract the position-independent quasinormal oscillation frequency $f_{\rm{peak}}(m)$ (see the appendix for details on data analysis).

\begin{figure}[!htp]
\includegraphics[trim = 0.5cm 0 0 0, scale=0.8]{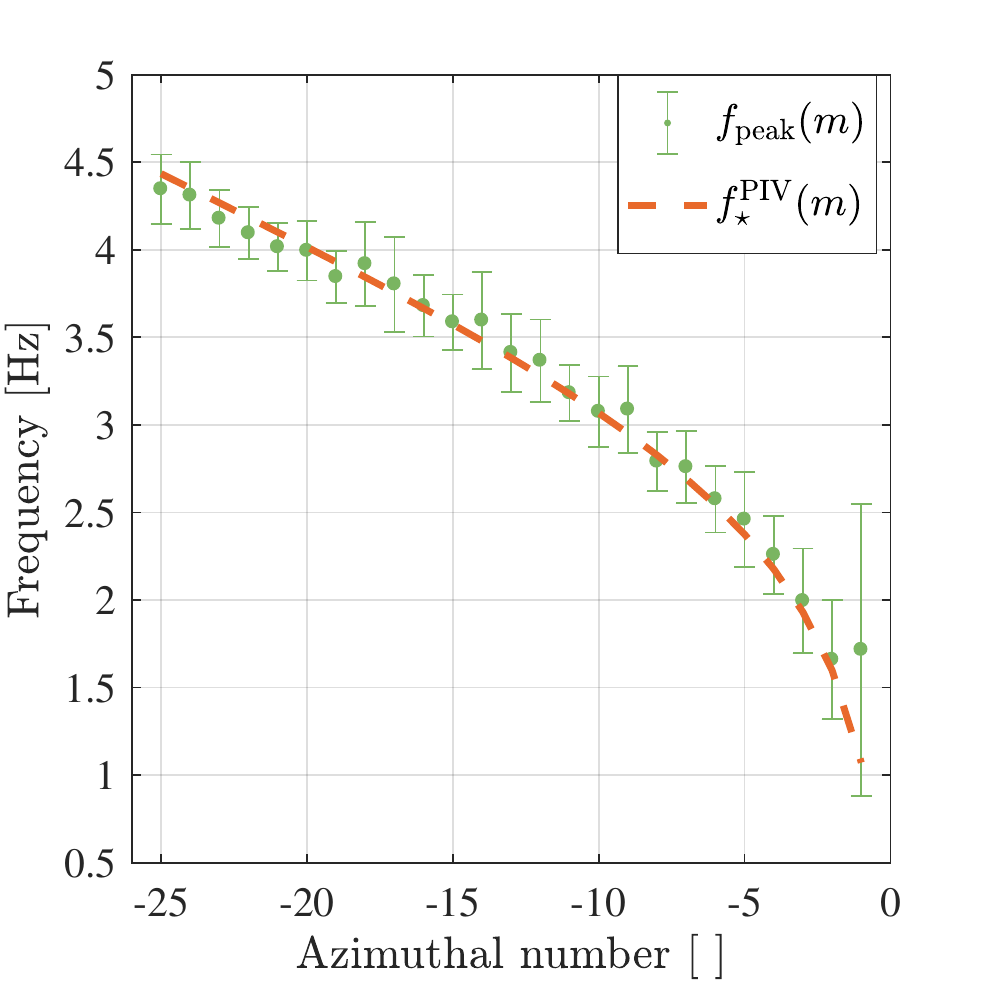}  
\caption{{\bf Characteristic spectrum of the Unruh vortex}. 
The quasinormal frequency spectrum $f_{\mathrm{peak}}(m)$, extracted from the experimental data and represented by green dots, is compared with the theoretical prediction for the light-ring frequencies, $f_{\star}^{\rm{PIV}}(m)$, computed for $C = 151~\rm{cm}^2/\rm{s}$ and $D = 0~\rm{cm}^2/\rm{s}$. These flow parameters were obtained via Particle Imaging Velocimetry (PIV). The error bars indicate the standard deviation over 25 experiments. The two spectra agree, confirming the detection of light-ring mode oscillations. 
}
\label{spec}
\end{figure}

In Fig.~\ref{spec} we present the quasinormal oscillation frequencies $f_{\rm{peak}}(m)$ extracted from the experimental data together with the predicted spectrum, $f_{\star}^{\rm{PIV}}(m)$, computed from Hamilton's equations \eqref{LRcond} using the flow parameters $C= 151~\rm{cm}^2/\rm{s}$ and $D = 0~\rm{cm}^2/\rm{s}$ obtained through PIV.
We observe that the LR model describing the characteristic oscillations of an Unruh vortex as quasinormal modes is consistent with the data. This is the first experimental observation of the oscillatory part of the quasinormal mode spectrum in an analogue black hole experiment.

\noindent {\bf \emph Final remarks.} Our experiment exhibits a new facet of the fluid-gravity analogy~\cite{Unruh,Visser93,Schutzhold} which has led to a better understanding of fundamental phenomena such as Hawking radiation~\cite{jacobson91,Unruh94} and superradiance~\cite{basak,basak2,torres2020estimate}.
It provides the first observation of quasinormal mode oscillations in analogue gravity and will, hopefully, pave the way for real-life applications of the fluid-gravity analogy. 

In particular, we would like to lay the emphasis on the fact that the first 25 counter-rotating modes have been detected and were simultaneously excited during this out-of-equilibrium process. In fact, the limitation for observing higher $m$-modes is purely technical and due to the resolution of the detection method used. We further recall that no external perturbations, such as incident waves, were added to the flow and that the characteristic modes were naturally emitted by the \textit{Unruh vortex}. Our results suggest that QNMs can easily be stimulated in laboratory systems, \emph{independently} of the $m$-value, and as such provide a novel and practical tool to study wave-current interactions.

For astrophysical black holes, for instance, the detection of quasinormal ringing through gravitational wave detectors~\cite{LIGO_GW,LIGO_propBH} opens up the possibility of \textit{Black Hole Spectroscopy}, i.e.~the identification of the spacetime geometry through the measurement of the frequency spectrum of gravitational waves emitted by a newly formed black hole~\cite{press,Echeverria,schutz}. For fluid flows, our succesful measurement of quasinormal oscillation frequencies opens up the possibility of fluid flow spectroscopy, as explained in~\cite{Torres_2019}. In principle, this method can be used as an alternative to the standard fluid flow visualisation techniques that require tracer particles. In particular, when suitable tracer particles are hardly found or do not exist, like in superfluids~\cite{Chopra}, this could provide a promising non-invasive method to characterise fluid flows.

Finally, we believe that our present result will motivate similar studies in fluids of light systems capable of establishing rotating geometries~\cite{Vocke:2017tif,Jacquet:2020znqf}.

\appendix

\section{Particle Imaging Velocimetry} \label{PIV_method}

To characterise the background flow we use Particle Imaging Velocimetry (PIV). PIV is a technique frequently used in engineering to measure fluid configurations. It relies on seeding the flow with small tracer particles and recording them using a high-speed camera as they follow the fluid streamlines. Each image thus obtained is split into smaller windows, which are then compared with similar windows in the consecutive image. For each window in a given image we compute the correlation table with respect to windows in the next image. By finding the maxima of the correlation tables, one is able to identify the displacement of each window between images. Repeating the procedure for all pairs of consecutive images, one can reconstruct the evolution of the velocity field in the region captured by the images (see e.g.~\cite{book_PIV}
for a thorough review).

In our experiment, the flow was seeded with plyolite particles with an average diameter of $1~{\rm mm}$. 
The particles were illuminated with a $2~{\rm mm}$-thick light sheet provided by a Yb-doped laser of characteristic wavelength $523~{\rm nm}$.
The laser was aligned in all cases to be pointing directly at the vortex and the height of the sheet was adjusted to $5.4~{\rm cm}$, i.e.~just below the water surface. 
Due to the presence of a shadow in the camera's field of view (located where the laser sheet intersects the vortex core), the laser source was placed at three different positions: half way along the length of the tank and offset by $1.2~{\rm m}$ to either side.
The camera was positioned such that the pixel size was $0.32~{\rm mm}$. 
Two measurements were taken for each laser position, giving a total of 6 experimental runs. 
Since PIV requires the presence of tracer particles, it was not possible to run PIV simultaneously with the Fast-Chequerboard Demodulation method.

In each experimental run the particles were recorded for $5.445{\rm s}$ using a Phantom Miro Lab 340 camera, with a Sigma 24-70mm F2.8 lens attached, at a resolution of 1600x1600 pixels and a frame rate of 200 frames per second (corresponding to 1090 images). The images were analysed using the MATLAB extension PIVlab, developed in~\cite{book_PIV}.
We used PIVlab's window deformation tool, with spline deformation, to reduce the number of erroneous vector identifications. 
The analysis was performed over three iterations using an interrogation area of 256 x 256, 128 x 128 and 64 x 64 pixels respectively, each with $50\%$
overlapping. 

\begin{figure}[!htp]
\includegraphics[scale = 0.7]{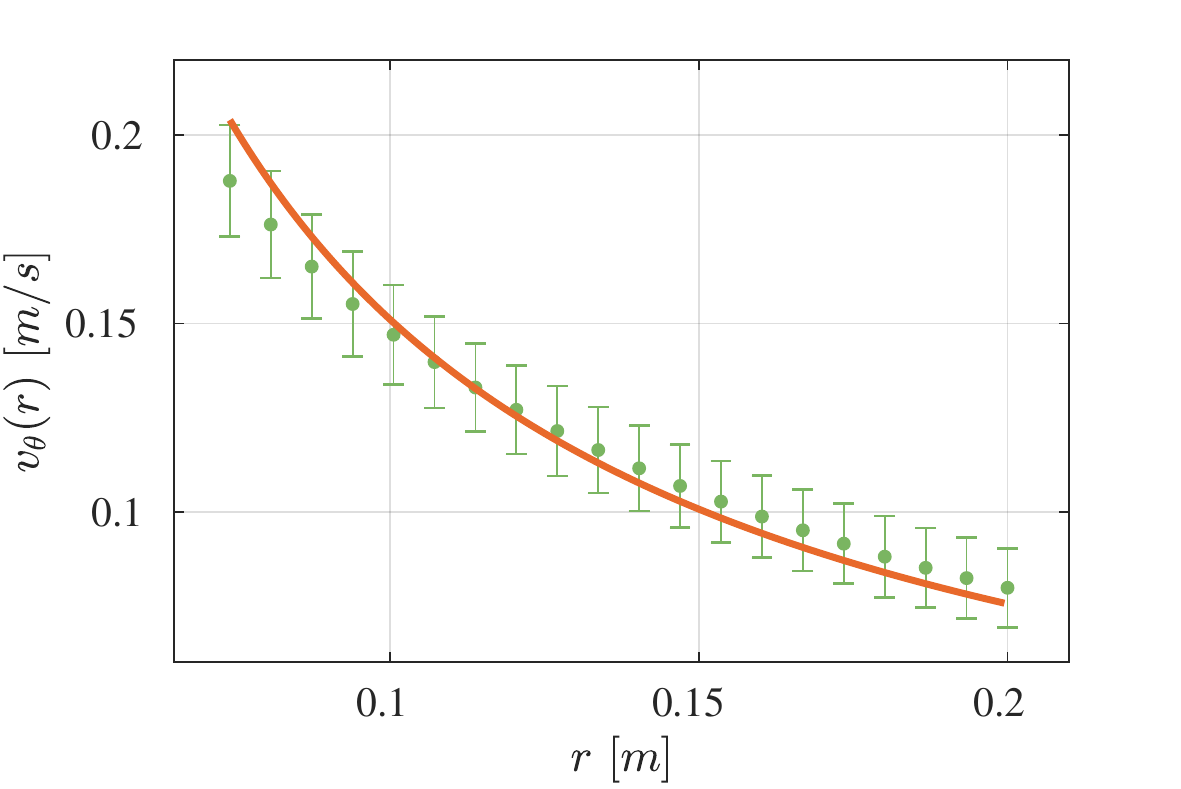} %
\caption{{\bf Angular component of the velocity profile}. Green dots correpond to the averaged (over time, angle and experiments) angular velocity. The error bars indicate the standard deviation. The orange curve is the weighted least-squares fit to the angular velocity $v^{\theta} = C/r$ of the Draining Bathtub model, corresponding to $C = 151$ $\mathrm{cm}^2/\mathrm{s}$.}
\end{figure}

The velocity field $\vec{v}(t_k,r_i,\theta_j)$ thus obtained is first averaged in time and then decomposed into an angular part, $v^{\theta}(r_i,\theta_j)$, and a radial one, $v^{r}(r_i,\theta_j)$, where $r_i \in  [7.4~\mathrm{cm}, 20~\mathrm{cm}]$. We note that $v^{r}\ll v^\theta$ in the region of observation. The maximum value of the radial velocity is approximately a tenth of the angular one. More importantly, the bias in our PIV method (estimated by applying the method to simulated data) has the same order of magnitude as the measured radial velocities. This implies that we cannot extract a reliable value for the drain parameter $D$ in the analysed region. In fact, it is expected that most of the draining in free surface vortices occurs through the bulk and through the boundary layer at the bottom of the tank, with a negligible radial flow at the surface, especially far from the vortex core~\cite{andersen1,andersen2,CRISTOFANO}.
 Nevertheless, we can determine an upper-bound for $D$, namely \mbox{$D_{\max} = \underset{i,j}{\max} [r_i v^r(r_i,\theta_j)] = 39~\rm{cm}^2/\rm{s}$}.

In the window of observation, $v^\theta(r_i,\theta_k)$ is found to be $\theta$-independent to a good approximation. This allows us to average the angular profile over the angles to obtain $v^\theta(r_i)$. To justify this procedure we make the azimuthal decomposition \mbox{$v^{\theta}(r_i,\theta_j) = v^{\theta}(r_i) + U_{\mathrm{asym}}(r_i,\theta_j)$}, where $U_{\mathrm{asym}}(r_i,\theta_j) = \underset{m\neq0}{\sum} v^{\theta,m}(r_i)e^{-im\theta_j}$ is the asymmetric part of the flow field. 
We then compare the energy of the asymmetric and axisymmetric parts, finding that
$\underset{m\neq0}{\sum} \mathcal{E}_m / \mathcal{E}_0 \sim 1.3\%$,
 where $\mathcal{E}_m = \underset{i}{\sum} |v^{\theta,m}(r_i)|^2$ is the energy of each azimuthal component. This indicates  that our flow is highly symmetric.

The averaged (over time, angle and experiments) angular velocity profile is displayed in Supplementary Fig.~1. The error associated with the averaging is given by the standard deviation and indicates the spread in the data about the mean value. 
To check the stationarity of the flow, we compare the angular velocity vector field in the first and final seconds of our experimental runs, finding that the maximum difference at any point is smaller than the uncertainty due to the vector identification inherent in the PIV. Hence, we deem the velocity field to be stationary within the error of our method.
Since a draining vortex is expected to be irrotational sufficiently far from the drain hole, we fit the mean angular velocity with the function $v^\theta(r) = C/r$, as shown in Supplementary Fig.~1. The extracted value for $C$ is $151~\rm{cm}^2/\rm{s}$.

\section{Data analysis.} \label{data_analysis}

In the following we outline the details of the detection method used and the various steps behind the data analysis leading to the \textit{Power Spectral Densities} (PSDs) shown in Fig.~1, to the radius-averaged PSDs shown in Fig.~2, and to the quasinormal frequency spectrum presented in Fig.~3. The free surface of the water is obtained using the Fast-Chequerboard Demodulation method~\cite{Sanders}.
 A periodic pattern is placed at the bottom of the tank in a region of \mbox{$59~ \mathrm{cm} \times 84~ \mathrm{cm}$}. 
The pattern is composed of two orthogonal sinusoidal waves with wavelengths of $6.5~\mathrm{mm}$ each.
Deformations of this pattern due to free surface fluctuations are recorded in a region of \mbox{$58~ \mathrm{cm} \times 58~ \mathrm{cm}$} over the vortex using a Phantom Camera Miro Lab 340 high speed camera at a frame rate of 40 $\mathrm{fps}$ over $16.3~ \mathrm{s}$ with an exposure time of 24000 $\mathrm{\mu s}$.
For each of the 651 pictures of the deformed pattern, we reconstruct the free surface in the form of an array $h(t_k,x_i,y_j)$ giving the height of the water at the $1600 \times 1600$ points on the free surface $(x_i, y_j)$ at every time step $t_k$\footnote{The MATLAB code used for this is available at: https://github.com/swildeman/fcd}. The typical amplitude of the free surface deformations corresponds, at most, to 2\% of the unperturbed water height. This justifies the linear treatment employed.

To select specific azimuthal numbers, we choose the centre of our coordinate system to be the centre of the hole and convert the signal from cartesian to polar coordinates. In addition to this change of coordinates, we discard all data points within a minimal radius $r_{\rm{min}}\approx 7.4~\mathrm{cm}$. 
This cropping is necessary as there is no clear pattern in the centre. 
This is either due to the hole at the bottom of the tank or due to the curvature of the vortex itself deforming the pattern too much to be detectable. 
We also discard points above a radius $r_{\rm{max}} \approx 25~\mathrm{cm}$, to avoid errors coming from the edge of the images.
After this step, our data is in the form $h(t_k,r_i,\theta_j)$ with   $r_{\rm{min}} < r_i < r_{\rm{max}}$. However, before selecting azimuthal components, we need to construct the analytic representation of the water elevation by adding to the real signal $h$ the imaginary part obtained from its Hilbert transform: \mbox{$h_\mathbb{C} = h + i \mathcal{H}(h)$}. 

The Hilbert transform $\mathcal{H}(h)$ is computed by means of a discrete Fourier transform and by removing the redundant negative frequency components of the time-spectrum. We have verified that other methods of computing the analytic representation, e.g.~wavelet transforms, give identical results. 
This step is crucial as it will allow us to distinguish between $m > 0$ and $m < 0$, i.e.~between waves that are co- and counter rotating with respect to the flow. 
At this stage we are left with a complex valued array $h_{\mathbb{C}}(t_k,r_i,\theta_j)$, such that $\mathrm{Re}(h_\mathbb{C}) = h$. In the following we will discard the index $\mathbb{C}$ and keep in mind that we are now dealing with a complex array.

We then perform a discrete Fourier transform in the angular direction to separate the various azimuthal components $h(t_k,r_i,m) = \sum_j h(t_k,r_i,\theta_j) e^{-im\theta_j} \Delta \theta$. 
From this, it is possible to extract the PSD of the waves emitted by the vortex for every azimuthal number m and every radius $r_i$,
\begin{equation}
PSD(f,r_i,m) \propto \left| \tilde{h}(f,r_i,m) \right|^2,
\end{equation}
where $\tilde{h}$ is the time Fourier transform of the height field at fixed $(m,r_i)$. In Fig.~1, we show the normalised PSDs for a range of $m$ values. The PSDs are finally averaged over the radius in order to look at the r-independent frequency content, i.e.~the oscillation frequency of the LR modes. Various averaged PSDs, denoted by $\mathcal{P}_m(f)$, are presented in Fig.~2. For each one of them, corresponding to a different $m$, the location of the peak, $f_{\rm{peak}}(m)$, is obtained by a parabolic interpolation of the maximum of the averaged PSD and its nearest neighbouring points. 

By repeating the procedure using a different choice for the center, located 10 pixels ($\approx 4\mathrm{mm}$) away from the original, we observed that the maximum deviation in the location of a frequency peak is approximately $2\%$, attesting the robustness of the procedure against an inaccurate choice for the centre.

\acknowledgments

We are indebted to the technical and administrative staff in the School of Physics \& Astronomy where our experimental setup is hosted. In particular, we want to thank Terry Wright, Tommy Napier and Ian Taylor for their support, hard work and sharing their technical knowledge and expertise with us to set up the experiment in Nottingham.
Furthermore we would like to thank Antonin Coutant, Sam Dolan, Sebastian Erne, Zack Fifer, Harry Goodhew, Cisco Gooding, J\"org Schmiedmayer, Ralf Sch\"utzhold, Thomas Sotiriou, and Bill Unruh for the many discussions on all aspects of the project. For comments on the paper, we wish to thank Michael Berry, Sebastian Erne, Juan Garrahan, Kostas Kokkotas, Luis Lehner, Igor Lesanovsky and Bill Unruh. 

M.~R.~acknowledges financial support from the S\~ao Paulo Research Foundation (FAPESP, Brazil), Grants No.
2013/09357-9 and 2018/10597-8, and from Conselho Nacional de Desenvolvimento Cient\'{i}fico e Tecnol\'{o}gico (CNPq, Brazil), Grant FA 309749/2017-4. M.~R.~is also grateful to the University of Nottingham for hospitality while this work was being completed. SW acknowledges financial
support provided under the Paper Enhancement Grant at
the University of Nottingham, the Royal Society University
Research Fellow (UF120112), the Nottingham Advanced
Research Fellow (A2RHS2), the Royal Society
Project (RG130377) grants, the Royal Society Enhancement
Grant (RGF/EA/180286) and the EPSRC Project
Grant (EP/P00637X/1). SW acknowledges partial support
from STFC consolidated grant No. ST/P000703/.

\bibliography{Abhs_prl_main_with_supp}

\begin{thebibliography}{42}%
\makeatletter
\providecommand \@ifxundefined [1]{%
 \@ifx{#1\undefined}
}%
\providecommand \@ifnum [1]{%
 \ifnum #1\expandafter \@firstoftwo
 \else \expandafter \@secondoftwo
 \fi
}%
\providecommand \@ifx [1]{%
 \ifx #1\expandafter \@firstoftwo
 \else \expandafter \@secondoftwo
 \fi
}%
\providecommand \natexlab [1]{#1}%
\providecommand \enquote  [1]{``#1''}%
\providecommand \bibnamefont  [1]{#1}%
\providecommand \bibfnamefont [1]{#1}%
\providecommand \citenamefont [1]{#1}%
\providecommand \href@noop [0]{\@secondoftwo}%
\providecommand \href [0]{\begingroup \@sanitize@url \@href}%
\providecommand \@href[1]{\@@startlink{#1}\@@href}%
\providecommand \@@href[1]{\endgroup#1\@@endlink}%
\providecommand \@sanitize@url [0]{\catcode `\\12\catcode `\$12\catcode
  `\&12\catcode `\#12\catcode `\^12\catcode `\_12\catcode `\%12\relax}%
\providecommand \@@startlink[1]{}%
\providecommand \@@endlink[0]{}%
\providecommand \url  [0]{\begingroup\@sanitize@url \@url }%
\providecommand \@url [1]{\endgroup\@href {#1}{\urlprefix }}%
\providecommand \urlprefix  [0]{URL }%
\providecommand \Eprint [0]{\href }%
\providecommand \doibase [0]{https://doi.org/}%
\providecommand \selectlanguage [0]{\@gobble}%
\providecommand \bibinfo  [0]{\@secondoftwo}%
\providecommand \bibfield  [0]{\@secondoftwo}%
\providecommand \translation [1]{[#1]}%
\providecommand \BibitemOpen [0]{}%
\providecommand \bibitemStop [0]{}%
\providecommand \bibitemNoStop [0]{.\EOS\space}%
\providecommand \EOS [0]{\spacefactor3000\relax}%
\providecommand \BibitemShut  [1]{\csname bibitem#1\endcsname}%
\let\auto@bib@innerbib\@empty
\bibitem [{\citenamefont {Barcel{\'o}}\ \emph {et~al.}(2011)\citenamefont
  {Barcel{\'o}}, \citenamefont {Liberati},\ and\ \citenamefont
  {Visser}}]{Barcelo05}%
  \BibitemOpen
  \bibfield  {author} {\bibinfo {author} {\bibfnamefont {C.}~\bibnamefont
  {Barcel{\'o}}}, \bibinfo {author} {\bibfnamefont {S.}~\bibnamefont
  {Liberati}},\ and\ \bibinfo {author} {\bibfnamefont {M.}~\bibnamefont
  {Visser}},\ }\href {https://doi.org/10.12942/lrr-2011-3} {\bibfield
  {journal} {\bibinfo  {journal} {Living Reviews in Relativity}\ }\textbf
  {\bibinfo {volume} {14}},\ \bibinfo {pages} {3} (\bibinfo {year}
  {2011})}\BibitemShut {NoStop}%
\bibitem [{\citenamefont {Barcel{\'o}}(2019)}]{Barcelo:2018ynq}%
  \BibitemOpen
  \bibfield  {author} {\bibinfo {author} {\bibfnamefont {C.}~\bibnamefont
  {Barcel{\'o}}},\ }\href {https://doi.org/10.1038/s41567-018-0367-6}
  {\bibfield  {journal} {\bibinfo  {journal} {Nature Phys.}\ }\textbf {\bibinfo
  {volume} {15}},\ \bibinfo {pages} {210} (\bibinfo {year} {2019})}\BibitemShut
  {NoStop}%
\bibitem [{\citenamefont {Weinfurtner}\ \emph {et~al.}(2011)\citenamefont
  {Weinfurtner}, \citenamefont {Tedford}, \citenamefont {Penrice},
  \citenamefont {Unruh},\ and\ \citenamefont {Lawrence}}]{Weinfurtner10}%
  \BibitemOpen
  \bibfield  {author} {\bibinfo {author} {\bibfnamefont {S.}~\bibnamefont
  {Weinfurtner}}, \bibinfo {author} {\bibfnamefont {E.~W.}\ \bibnamefont
  {Tedford}}, \bibinfo {author} {\bibfnamefont {M.~C.~J.}\ \bibnamefont
  {Penrice}}, \bibinfo {author} {\bibfnamefont {W.~G.}\ \bibnamefont {Unruh}},\
  and\ \bibinfo {author} {\bibfnamefont {G.~A.}\ \bibnamefont {Lawrence}},\
  }\href {https://doi.org/10.1103/PhysRevLett.106.021302} {\bibfield  {journal}
  {\bibinfo  {journal} {Phys. Rev. Lett.}\ }\textbf {\bibinfo {volume} {106}},\
  \bibinfo {pages} {021302} (\bibinfo {year} {2011})},\ \Eprint
  {https://arxiv.org/abs/1008.1911} {arXiv:1008.1911 [gr-qc]} \BibitemShut
  {NoStop}%
\bibitem [{\citenamefont {Steinhauer}(2016)}]{Steinhauer15}%
  \BibitemOpen
  \bibfield  {author} {\bibinfo {author} {\bibfnamefont {J.}~\bibnamefont
  {Steinhauer}},\ }\href {https://doi.org/10.1038/nphys3863} {\bibfield
  {journal} {\bibinfo  {journal} {Nature Phys.}\ }\textbf {\bibinfo {volume}
  {12}},\ \bibinfo {pages} {959} (\bibinfo {year} {2016})},\ \Eprint
  {https://arxiv.org/abs/1510.00621} {arXiv:1510.00621 [gr-qc]} \BibitemShut
  {NoStop}%
\bibitem [{\citenamefont {Euv\'e}\ \emph {et~al.}(2016)\citenamefont {Euv\'e},
  \citenamefont {Michel}, \citenamefont {Parentani}, \citenamefont {Philbin},\
  and\ \citenamefont {Rousseaux}}]{Euve16}%
  \BibitemOpen
  \bibfield  {author} {\bibinfo {author} {\bibfnamefont {L.-P.}\ \bibnamefont
  {Euv\'e}}, \bibinfo {author} {\bibfnamefont {F.}~\bibnamefont {Michel}},
  \bibinfo {author} {\bibfnamefont {R.}~\bibnamefont {Parentani}}, \bibinfo
  {author} {\bibfnamefont {T.~G.}\ \bibnamefont {Philbin}},\ and\ \bibinfo
  {author} {\bibfnamefont {G.}~\bibnamefont {Rousseaux}},\ }\href
  {https://doi.org/10.1103/PhysRevLett.117.121301} {\bibfield  {journal}
  {\bibinfo  {journal} {Phys. Rev. Lett.}\ }\textbf {\bibinfo {volume} {117}},\
  \bibinfo {pages} {121301} (\bibinfo {year} {2016})}\BibitemShut {NoStop}%
\bibitem [{\citenamefont {Mu{\~{n}}oz~de Nova}\ \emph
  {et~al.}(2019)\citenamefont {Mu{\~{n}}oz~de Nova}, \citenamefont {Golubkov},
  \citenamefont {Kolobov},\ and\ \citenamefont {Steinhauer}}]{deNova:2018rld}%
  \BibitemOpen
  \bibfield  {author} {\bibinfo {author} {\bibfnamefont {J.~R.}\ \bibnamefont
  {Mu{\~{n}}oz~de Nova}}, \bibinfo {author} {\bibfnamefont {K.}~\bibnamefont
  {Golubkov}}, \bibinfo {author} {\bibfnamefont {V.~I.}\ \bibnamefont
  {Kolobov}},\ and\ \bibinfo {author} {\bibfnamefont {J.}~\bibnamefont
  {Steinhauer}},\ }\href {https://doi.org/10.1038/s41586-019-1241-0} {\bibfield
   {journal} {\bibinfo  {journal} {Nature}\ }\textbf {\bibinfo {volume}
  {569}},\ \bibinfo {pages} {688} (\bibinfo {year} {2019})},\ \Eprint
  {https://arxiv.org/abs/1809.00913} {arXiv:1809.00913 [gr-qc]} \BibitemShut
  {NoStop}%
\bibitem [{\citenamefont {Torres}\ \emph {et~al.}(2017)\citenamefont {Torres},
  \citenamefont {Patrick}, \citenamefont {Coutant}, \citenamefont {Richartz},
  \citenamefont {Tedford},\ and\ \citenamefont {Weinfurtner}}]{Torres16}%
  \BibitemOpen
  \bibfield  {author} {\bibinfo {author} {\bibfnamefont {T.}~\bibnamefont
  {Torres}}, \bibinfo {author} {\bibfnamefont {S.}~\bibnamefont {Patrick}},
  \bibinfo {author} {\bibfnamefont {A.}~\bibnamefont {Coutant}}, \bibinfo
  {author} {\bibfnamefont {M.}~\bibnamefont {Richartz}}, \bibinfo {author}
  {\bibfnamefont {E.~W.}\ \bibnamefont {Tedford}},\ and\ \bibinfo {author}
  {\bibfnamefont {S.}~\bibnamefont {Weinfurtner}},\ }\href
  {https://doi.org/10.1038/nphys4151} {\bibfield  {journal} {\bibinfo
  {journal} {Nature Phys.}\ }\textbf {\bibinfo {volume} {13}},\ \bibinfo
  {pages} {833} (\bibinfo {year} {2017})},\ \Eprint
  {https://arxiv.org/abs/1612.06180} {arXiv:1612.06180 [gr-qc]} \BibitemShut
  {NoStop}%
\bibitem [{\citenamefont {Unruh}(1981)}]{Unruh}%
  \BibitemOpen
  \bibfield  {author} {\bibinfo {author} {\bibfnamefont {W.}~\bibnamefont
  {Unruh}},\ }\href {https://doi.org/10.1103/PhysRevLett.46.1351} {\bibfield
  {journal} {\bibinfo  {journal} {Phys. Rev. Lett.}\ }\textbf {\bibinfo
  {volume} {46}},\ \bibinfo {pages} {1351} (\bibinfo {year}
  {1981})}\BibitemShut {NoStop}%
\bibitem [{\citenamefont {Visser}(1993)}]{Visser93}%
  \BibitemOpen
  \bibfield  {author} {\bibinfo {author} {\bibfnamefont {M.}~\bibnamefont
  {Visser}},\ }\href@noop {} {\bibfield  {journal} {\bibinfo  {journal} {ArXiv:
  gr-qc/9311028}\ } (\bibinfo {year} {1993})},\ \Eprint
  {https://arxiv.org/abs/gr-qc/9311028} {arXiv:gr-qc/9311028 [gr-qc]}
  \BibitemShut {NoStop}%
\bibitem [{\citenamefont {{Sch{\"u}tzhold}}\ and\ \citenamefont
  {{Unruh}}(2002)}]{Schutzhold}%
  \BibitemOpen
  \bibfield  {author} {\bibinfo {author} {\bibfnamefont {R.}~\bibnamefont
  {{Sch{\"u}tzhold}}}\ and\ \bibinfo {author} {\bibfnamefont {W.~G.}\
  \bibnamefont {{Unruh}}},\ }\href {https://doi.org/10.1103/PhysRevD.66.044019}
  {\bibfield  {journal} {\bibinfo  {journal} {Phys. Rev. D}\ }\textbf {\bibinfo
  {volume} {66}},\ \bibinfo {eid} {044019} (\bibinfo {year}
  {2002})}\BibitemShut {NoStop}%
\bibitem [{\citenamefont {Kokkotas}\ and\ \citenamefont
  {Schmidt}(1999)}]{Kokkotas99}%
  \BibitemOpen
  \bibfield  {author} {\bibinfo {author} {\bibfnamefont {K.~D.}\ \bibnamefont
  {Kokkotas}}\ and\ \bibinfo {author} {\bibfnamefont {B.~G.}\ \bibnamefont
  {Schmidt}},\ }\href {https://doi.org/10.12942/lrr-1999-2} {\bibfield
  {journal} {\bibinfo  {journal} {Living Rev. Rel.}\ }\textbf {\bibinfo
  {volume} {2}},\ \bibinfo {pages} {2} (\bibinfo {year} {1999})},\ \Eprint
  {https://arxiv.org/abs/gr-qc/9909058} {arXiv:gr-qc/9909058 [gr-qc]}
  \BibitemShut {NoStop}%
\bibitem [{\citenamefont {Berti}\ \emph {et~al.}(2009)\citenamefont {Berti},
  \citenamefont {Cardoso},\ and\ \citenamefont {Starinets}}]{Berti}%
  \BibitemOpen
  \bibfield  {author} {\bibinfo {author} {\bibfnamefont {E.}~\bibnamefont
  {Berti}}, \bibinfo {author} {\bibfnamefont {V.}~\bibnamefont {Cardoso}},\
  and\ \bibinfo {author} {\bibfnamefont {A.~O.}\ \bibnamefont {Starinets}},\
  }\href {https://doi.org/10.1088/0264-9381/26/16/163001} {\bibfield  {journal}
  {\bibinfo  {journal} {Class. Quant. Grav.}\ }\textbf {\bibinfo {volume}
  {26}},\ \bibinfo {pages} {163001} (\bibinfo {year} {2009})},\ \Eprint
  {https://arxiv.org/abs/0905.2975} {arXiv:0905.2975 [gr-qc]} \BibitemShut
  {NoStop}%
\bibitem [{\citenamefont {Konoplya}\ and\ \citenamefont
  {Zhidenko}(2011)}]{Konoplya11}%
  \BibitemOpen
  \bibfield  {author} {\bibinfo {author} {\bibfnamefont {R.~A.}\ \bibnamefont
  {Konoplya}}\ and\ \bibinfo {author} {\bibfnamefont {A.}~\bibnamefont
  {Zhidenko}},\ }\href {https://doi.org/10.1103/RevModPhys.83.793} {\bibfield
  {journal} {\bibinfo  {journal} {Rev. Mod. Phys.}\ }\textbf {\bibinfo {volume}
  {83}},\ \bibinfo {pages} {793} (\bibinfo {year} {2011})},\ \Eprint
  {https://arxiv.org/abs/1102.4014} {arXiv:1102.4014 [gr-qc]} \BibitemShut
  {NoStop}%
\bibitem [{\citenamefont {Berti}\ \emph {et~al.}(2004)\citenamefont {Berti},
  \citenamefont {Cardoso},\ and\ \citenamefont {Lemos}}]{Berti:2004ju}%
  \BibitemOpen
  \bibfield  {author} {\bibinfo {author} {\bibfnamefont {E.}~\bibnamefont
  {Berti}}, \bibinfo {author} {\bibfnamefont {V.}~\bibnamefont {Cardoso}},\
  and\ \bibinfo {author} {\bibfnamefont {J.~P.~S.}\ \bibnamefont {Lemos}},\
  }\href {https://doi.org/10.1103/PhysRevD.70.124006} {\bibfield  {journal}
  {\bibinfo  {journal} {Phys. Rev.}\ }\textbf {\bibinfo {volume} {D70}},\
  \bibinfo {pages} {124006} (\bibinfo {year} {2004})},\ \Eprint
  {https://arxiv.org/abs/gr-qc/0408099} {arXiv:gr-qc/0408099 [gr-qc]}
  \BibitemShut {NoStop}%
\bibitem [{\citenamefont {Dolan}\ \emph {et~al.}(2012)\citenamefont {Dolan},
  \citenamefont {Oliveira},\ and\ \citenamefont {Crispino}}]{Dolan3}%
  \BibitemOpen
  \bibfield  {author} {\bibinfo {author} {\bibfnamefont {S.~R.}\ \bibnamefont
  {Dolan}}, \bibinfo {author} {\bibfnamefont {L.~A.}\ \bibnamefont
  {Oliveira}},\ and\ \bibinfo {author} {\bibfnamefont {L.~C.~B.}\ \bibnamefont
  {Crispino}},\ }\href {https://doi.org/10.1103/PhysRevD.85.044031} {\bibfield
  {journal} {\bibinfo  {journal} {Phys. Rev.}\ }\textbf {\bibinfo {volume} {D
  85}},\ \bibinfo {pages} {044031} (\bibinfo {year} {2012})},\ \Eprint
  {https://arxiv.org/abs/1105.1795} {arXiv:1105.1795 [gr-qc]} \BibitemShut
  {NoStop}%
\bibitem [{\citenamefont {Patrick}\ \emph {et~al.}(2018)\citenamefont
  {Patrick}, \citenamefont {Coutant}, \citenamefont {Richartz},\ and\
  \citenamefont {Weinfurtner}}]{flowmaster}%
  \BibitemOpen
  \bibfield  {author} {\bibinfo {author} {\bibfnamefont {S.}~\bibnamefont
  {Patrick}}, \bibinfo {author} {\bibfnamefont {A.}~\bibnamefont {Coutant}},
  \bibinfo {author} {\bibfnamefont {M.}~\bibnamefont {Richartz}},\ and\
  \bibinfo {author} {\bibfnamefont {S.}~\bibnamefont {Weinfurtner}},\ }\href
  {https://doi.org/10.1103/PhysRevLett.121.061101} {\bibfield  {journal}
  {\bibinfo  {journal} {Phys. Rev. Lett.}\ }\textbf {\bibinfo {volume} {121}},\
  \bibinfo {pages} {061101} (\bibinfo {year} {2018})},\ \Eprint
  {https://arxiv.org/abs/1801.08473} {arXiv:1801.08473 [gr-qc]} \BibitemShut
  {NoStop}%
\bibitem [{\citenamefont {Torres}\ \emph {et~al.}(2018)\citenamefont {Torres},
  \citenamefont {Coutant}, \citenamefont {Dolan},\ and\ \citenamefont
  {Weinfurtner}}]{TheoLR}%
  \BibitemOpen
  \bibfield  {author} {\bibinfo {author} {\bibfnamefont {T.}~\bibnamefont
  {Torres}}, \bibinfo {author} {\bibfnamefont {A.}~\bibnamefont {Coutant}},
  \bibinfo {author} {\bibfnamefont {S.}~\bibnamefont {Dolan}},\ and\ \bibinfo
  {author} {\bibfnamefont {S.}~\bibnamefont {Weinfurtner}},\ }\href
  {https://doi.org/10.1017/jfm.2018.752} {\bibfield  {journal} {\bibinfo
  {journal} {J. Fluid Mech.}\ }\textbf {\bibinfo {volume} {857}},\ \bibinfo
  {pages} {291} (\bibinfo {year} {2018})},\ \Eprint
  {https://arxiv.org/abs/1712.04675} {arXiv:1712.04675 [gr-qc]} \BibitemShut
  {NoStop}%
\bibitem [{\citenamefont {{Goebel}}(1972)}]{Goebel72}%
  \BibitemOpen
  \bibfield  {author} {\bibinfo {author} {\bibfnamefont {C.~J.}\ \bibnamefont
  {{Goebel}}},\ }\href {https://doi.org/10.1086/180898} {\bibfield  {journal}
  {\bibinfo  {journal} {Astrophys. J.}\ }\textbf {\bibinfo {volume} {172}},\
  \bibinfo {pages} {L95} (\bibinfo {year} {1972})}\BibitemShut {NoStop}%
\bibitem [{\citenamefont {Cardoso}\ \emph {et~al.}(2009)\citenamefont
  {Cardoso}, \citenamefont {Miranda}, \citenamefont {Berti}, \citenamefont
  {Witek},\ and\ \citenamefont {Zanchin}}]{Cardoso}%
  \BibitemOpen
  \bibfield  {author} {\bibinfo {author} {\bibfnamefont {V.}~\bibnamefont
  {Cardoso}}, \bibinfo {author} {\bibfnamefont {A.~S.}\ \bibnamefont
  {Miranda}}, \bibinfo {author} {\bibfnamefont {E.}~\bibnamefont {Berti}},
  \bibinfo {author} {\bibfnamefont {H.}~\bibnamefont {Witek}},\ and\ \bibinfo
  {author} {\bibfnamefont {V.~T.}\ \bibnamefont {Zanchin}},\ }\href
  {https://doi.org/10.1103/PhysRevD.79.064016} {\bibfield  {journal} {\bibinfo
  {journal} {Phys. Rev. D}\ }\textbf {\bibinfo {volume} {79}},\ \bibinfo
  {pages} {064016} (\bibinfo {year} {2009})}\BibitemShut {NoStop}%
\bibitem [{\citenamefont {Khanna}\ and\ \citenamefont {Price}(2017)}]{khanna}%
  \BibitemOpen
  \bibfield  {author} {\bibinfo {author} {\bibfnamefont {G.}~\bibnamefont
  {Khanna}}\ and\ \bibinfo {author} {\bibfnamefont {R.~H.}\ \bibnamefont
  {Price}},\ }\href {https://doi.org/10.1103/PhysRevD.95.081501} {\bibfield
  {journal} {\bibinfo  {journal} {Phys. Rev.}\ }\textbf {\bibinfo {volume}
  {D95}},\ \bibinfo {pages} {081501} (\bibinfo {year} {2017})},\ \Eprint
  {https://arxiv.org/abs/1609.00083} {arXiv:1609.00083 [gr-qc]} \BibitemShut
  {NoStop}%
\bibitem [{\citenamefont {Dolan}\ and\ \citenamefont
  {Oliveira}(2013)}]{Dolan2}%
  \BibitemOpen
  \bibfield  {author} {\bibinfo {author} {\bibfnamefont {S.~R.}\ \bibnamefont
  {Dolan}}\ and\ \bibinfo {author} {\bibfnamefont {E.~S.}\ \bibnamefont
  {Oliveira}},\ }\href {https://doi.org/10.1103/PhysRevD.87.124038} {\bibfield
  {journal} {\bibinfo  {journal} {Phys. Rev.}\ }\textbf {\bibinfo {volume}
  {D87}},\ \bibinfo {pages} {124038} (\bibinfo {year} {2013})}\BibitemShut
  {NoStop}%
\bibitem [{\citenamefont {Richartz}\ \emph {et~al.}(2015)\citenamefont
  {Richartz}, \citenamefont {Prain}, \citenamefont {Liberati},\ and\
  \citenamefont {Weinfurtner}}]{Mauricio14}%
  \BibitemOpen
  \bibfield  {author} {\bibinfo {author} {\bibfnamefont {M.}~\bibnamefont
  {Richartz}}, \bibinfo {author} {\bibfnamefont {A.}~\bibnamefont {Prain}},
  \bibinfo {author} {\bibfnamefont {S.}~\bibnamefont {Liberati}},\ and\
  \bibinfo {author} {\bibfnamefont {S.}~\bibnamefont {Weinfurtner}},\ }\href
  {https://doi.org/10.1103/PhysRevD.91.124018} {\bibfield  {journal} {\bibinfo
  {journal} {Phys. Rev.}\ }\textbf {\bibinfo {volume} {D91}},\ \bibinfo {pages}
  {124018} (\bibinfo {year} {2015})},\ \Eprint
  {https://arxiv.org/abs/1411.1662} {arXiv:1411.1662 [gr-qc]} \BibitemShut
  {NoStop}%
\bibitem [{\citenamefont {Nazarenko}(1994)}]{nazarenko}%
  \BibitemOpen
  \bibfield  {author} {\bibinfo {author} {\bibfnamefont {S.~V.}\ \bibnamefont
  {Nazarenko}},\ }\href {https://doi.org/10.1103/PhysRevLett.73.1793}
  {\bibfield  {journal} {\bibinfo  {journal} {Phys. Rev. Lett.}\ }\textbf
  {\bibinfo {volume} {73}},\ \bibinfo {pages} {1793} (\bibinfo {year}
  {1994})}\BibitemShut {NoStop}%
\bibitem [{\citenamefont {{Wildeman}}(2018)}]{Sanders}%
  \BibitemOpen
  \bibfield  {author} {\bibinfo {author} {\bibfnamefont {S.}~\bibnamefont
  {{Wildeman}}},\ }\href {https://doi.org/10.1007/s00348-018-2553-9} {\bibfield
   {journal} {\bibinfo  {journal} {Experiments in Fluids}\ }\textbf {\bibinfo
  {volume} {59}},\ \bibinfo {eid} {97} (\bibinfo {year} {2018})},\ \Eprint
  {https://arxiv.org/abs/1712.05679} {arXiv:1712.05679 [physics.flu-dyn]}
  \BibitemShut {NoStop}%
\bibitem [{\citenamefont {Jacobson}(1991)}]{jacobson91}%
  \BibitemOpen
  \bibfield  {author} {\bibinfo {author} {\bibfnamefont {T.}~\bibnamefont
  {Jacobson}},\ }\href {https://doi.org/10.1103/PhysRevD.44.1731} {\bibfield
  {journal} {\bibinfo  {journal} {Phys. Rev.}\ }\textbf {\bibinfo {volume}
  {D44}},\ \bibinfo {pages} {1731} (\bibinfo {year} {1991})}\BibitemShut
  {NoStop}%
\bibitem [{\citenamefont {Unruh}(1995)}]{Unruh94}%
  \BibitemOpen
  \bibfield  {author} {\bibinfo {author} {\bibfnamefont {W.~G.}\ \bibnamefont
  {Unruh}},\ }\href {https://doi.org/10.1103/PhysRevD.51.2827} {\bibfield
  {journal} {\bibinfo  {journal} {Phys. Rev.}\ }\textbf {\bibinfo {volume}
  {D51}},\ \bibinfo {pages} {2827} (\bibinfo {year} {1995})}\BibitemShut
  {NoStop}%
\bibitem [{\citenamefont {Basak}\ and\ \citenamefont
  {Majumdar}(2003{\natexlab{a}})}]{basak}%
  \BibitemOpen
  \bibfield  {author} {\bibinfo {author} {\bibfnamefont {S.}~\bibnamefont
  {Basak}}\ and\ \bibinfo {author} {\bibfnamefont {P.}~\bibnamefont
  {Majumdar}},\ }\href@noop {} {\bibfield  {journal} {\bibinfo  {journal}
  {Class. Quant. Grav.}\ }\textbf {\bibinfo {volume} {20}},\ \bibinfo {pages}
  {3907} (\bibinfo {year} {2003}{\natexlab{a}})}\BibitemShut {NoStop}%
\bibitem [{\citenamefont {Basak}\ and\ \citenamefont
  {Majumdar}(2003{\natexlab{b}})}]{basak2}%
  \BibitemOpen
  \bibfield  {author} {\bibinfo {author} {\bibfnamefont {S.}~\bibnamefont
  {Basak}}\ and\ \bibinfo {author} {\bibfnamefont {P.}~\bibnamefont
  {Majumdar}},\ }\href {https://doi.org/10.1088/0264-9381/20/13/335} {\bibfield
   {journal} {\bibinfo  {journal} {Class. Quant. Grav.}\ }\textbf {\bibinfo
  {volume} {20}},\ \bibinfo {pages} {2929} (\bibinfo {year}
  {2003}{\natexlab{b}})}\BibitemShut {NoStop}%
\bibitem [{\citenamefont {Torres}(2020)}]{torres2020estimate}%
  \BibitemOpen
  \bibfield  {author} {\bibinfo {author} {\bibfnamefont {T.}~\bibnamefont
  {Torres}},\ }\href@noop {} {\bibinfo {title} {Estimate of the superradiance
  spectrum in dispersive media}} (\bibinfo {year} {2020}),\ \Eprint
  {https://arxiv.org/abs/2003.02230} {arXiv:2003.02230 [gr-qc]} \BibitemShut
  {NoStop}%
\bibitem [{\citenamefont {Abbott}\ \emph
  {et~al.}(2016{\natexlab{a}})\citenamefont {Abbott} \emph {et~al.}}]{LIGO_GW}%
  \BibitemOpen
  \bibfield  {author} {\bibinfo {author} {\bibfnamefont {B.~P.}\ \bibnamefont
  {Abbott}} \emph {et~al.} (\bibinfo {collaboration} {Virgo, LIGO
  Scientific}),\ }\href@noop {} {\bibfield  {journal} {\bibinfo  {journal}
  {Phys. Rev. Lett.}\ }\textbf {\bibinfo {volume} {116}},\ \bibinfo {pages}
  {061102} (\bibinfo {year} {2016}{\natexlab{a}})},\ \Eprint
  {https://arxiv.org/abs/1602.03837} {arXiv:1602.03837 [gr-qc]} \BibitemShut
  {NoStop}%
\bibitem [{\citenamefont {Abbott}\ \emph
  {et~al.}(2016{\natexlab{b}})\citenamefont {Abbott} \emph
  {et~al.}}]{LIGO_propBH}%
  \BibitemOpen
  \bibfield  {author} {\bibinfo {author} {\bibfnamefont {B.~P.}\ \bibnamefont
  {Abbott}} \emph {et~al.} (\bibinfo {collaboration} {Virgo, LIGO
  Scientific}),\ }\href {https://doi.org/10.1103/PhysRevLett.116.241102}
  {\bibfield  {journal} {\bibinfo  {journal} {Phys. Rev. Lett.}\ }\textbf
  {\bibinfo {volume} {116}},\ \bibinfo {pages} {241102} (\bibinfo {year}
  {2016}{\natexlab{b}})},\ \Eprint {https://arxiv.org/abs/1602.03840}
  {arXiv:1602.03840 [gr-qc]} \BibitemShut {NoStop}%
\bibitem [{\citenamefont {Press}\ and\ \citenamefont {Thorne}(1972)}]{press}%
  \BibitemOpen
  \bibfield  {author} {\bibinfo {author} {\bibfnamefont {W.~H.}\ \bibnamefont
  {Press}}\ and\ \bibinfo {author} {\bibfnamefont {K.~S.}\ \bibnamefont
  {Thorne}},\ }\href {https://doi.org/10.1146/annurev.aa.10.090172.002003}
  {\bibfield  {journal} {\bibinfo  {journal} {Ann. Rev. Astron. Astrophys.}\
  }\textbf {\bibinfo {volume} {10}},\ \bibinfo {pages} {335} (\bibinfo {year}
  {1972})}\BibitemShut {NoStop}%
\bibitem [{\citenamefont {Echeverria}(1989)}]{Echeverria}%
  \BibitemOpen
  \bibfield  {author} {\bibinfo {author} {\bibfnamefont {F.}~\bibnamefont
  {Echeverria}},\ }\href {https://doi.org/10.1103/PhysRevD.40.3194} {\bibfield
  {journal} {\bibinfo  {journal} {Phys. Rev. D}\ }\textbf {\bibinfo {volume}
  {40}},\ \bibinfo {pages} {3194} (\bibinfo {year} {1989})}\BibitemShut
  {NoStop}%
\bibitem [{\citenamefont {Sathyaprakash}\ and\ \citenamefont
  {Schutz}(2009)}]{schutz}%
  \BibitemOpen
  \bibfield  {author} {\bibinfo {author} {\bibfnamefont {B.~S.}\ \bibnamefont
  {Sathyaprakash}}\ and\ \bibinfo {author} {\bibfnamefont {B.~F.}\ \bibnamefont
  {Schutz}},\ }\href {https://doi.org/10.12942/lrr-2009-2} {\bibfield
  {journal} {\bibinfo  {journal} {Living Reviews in Relativity}\ }\textbf
  {\bibinfo {volume} {12}},\ \bibinfo {pages} {2} (\bibinfo {year}
  {2009})}\BibitemShut {NoStop}%
\bibitem [{\citenamefont {Torres}\ \emph {et~al.}(2019)\citenamefont {Torres},
  \citenamefont {Patrick}, \citenamefont {Richartz},\ and\ \citenamefont
  {Weinfurtner}}]{Torres_2019}%
  \BibitemOpen
  \bibfield  {author} {\bibinfo {author} {\bibfnamefont {T.}~\bibnamefont
  {Torres}}, \bibinfo {author} {\bibfnamefont {S.}~\bibnamefont {Patrick}},
  \bibinfo {author} {\bibfnamefont {M.}~\bibnamefont {Richartz}},\ and\
  \bibinfo {author} {\bibfnamefont {S.}~\bibnamefont {Weinfurtner}},\ }\href
  {https://doi.org/10.1088/1361-6382/ab3d48} {\bibfield  {journal} {\bibinfo
  {journal} {Classical and Quantum Gravity}\ }\textbf {\bibinfo {volume}
  {36}},\ \bibinfo {pages} {194002} (\bibinfo {year} {2019})}\BibitemShut
  {NoStop}%
\bibitem [{\citenamefont {Chopra}\ and\ \citenamefont {Brown}(1957)}]{Chopra}%
  \BibitemOpen
  \bibfield  {author} {\bibinfo {author} {\bibfnamefont {K.~L.}\ \bibnamefont
  {Chopra}}\ and\ \bibinfo {author} {\bibfnamefont {J.~B.}\ \bibnamefont
  {Brown}},\ }\href {https://doi.org/10.1103/PhysRev.108.157} {\bibfield
  {journal} {\bibinfo  {journal} {Phys. Rev.}\ }\textbf {\bibinfo {volume}
  {108}},\ \bibinfo {pages} {157} (\bibinfo {year} {1957})}\BibitemShut
  {NoStop}%
\bibitem [{\citenamefont {Vocke}\ \emph {et~al.}(2017)\citenamefont {Vocke},
  \citenamefont {Maitland}, \citenamefont {Prain}, \citenamefont {Biancalana},
  \citenamefont {Marino},\ and\ \citenamefont {Faccio}}]{Vocke:2017tif}%
  \BibitemOpen
  \bibfield  {author} {\bibinfo {author} {\bibfnamefont {D.}~\bibnamefont
  {Vocke}}, \bibinfo {author} {\bibfnamefont {C.}~\bibnamefont {Maitland}},
  \bibinfo {author} {\bibfnamefont {A.}~\bibnamefont {Prain}}, \bibinfo
  {author} {\bibfnamefont {F.}~\bibnamefont {Biancalana}}, \bibinfo {author}
  {\bibfnamefont {F.}~\bibnamefont {Marino}},\ and\ \bibinfo {author}
  {\bibfnamefont {D.}~\bibnamefont {Faccio}},\ }\href@noop {} {\  (\bibinfo
  {year} {2017})},\ \Eprint {https://arxiv.org/abs/1709.04293}
  {arXiv:1709.04293 [physics.optics]} \BibitemShut {NoStop}%
\bibitem [{\citenamefont {Jacquet}\ \emph {et~al.}(2020)\citenamefont {Jacquet}
  \emph {et~al.}}]{Jacquet:2020znqf}%
  \BibitemOpen
  \bibfield  {author} {\bibinfo {author} {\bibfnamefont {M.~J.}\ \bibnamefont
  {Jacquet}} \emph {et~al.},\ }\href@noop {} {\  (\bibinfo {year} {2020})},\
  \Eprint {https://arxiv.org/abs/2002.00043} {arXiv:2002.00043
  [cond-mat.quant-gas]} \BibitemShut {NoStop}%
\bibitem [{\citenamefont {Raffel}\ \emph {et~al.}(2018)\citenamefont {Raffel},
  \citenamefont {Willert}, \citenamefont {Scarano}, \citenamefont {K{\"a}hler},
  \citenamefont {Wereley},\ and\ \citenamefont {Kompenhans}}]{book_PIV}%
  \BibitemOpen
  \bibfield  {author} {\bibinfo {author} {\bibfnamefont {M.}~\bibnamefont
  {Raffel}}, \bibinfo {author} {\bibfnamefont {C.}~\bibnamefont {Willert}},
  \bibinfo {author} {\bibfnamefont {F.}~\bibnamefont {Scarano}}, \bibinfo
  {author} {\bibfnamefont {C.}~\bibnamefont {K{\"a}hler}}, \bibinfo {author}
  {\bibfnamefont {S.}~\bibnamefont {Wereley}},\ and\ \bibinfo {author}
  {\bibfnamefont {J.}~\bibnamefont {Kompenhans}},\ }\href
  {https://books.google.co.uk/books?id=wk9UDwAAQBAJ} {\emph {\bibinfo {title}
  {Particle Image Velocimetry: A Practical Guide}}},\ Experimental Fluid
  Mechanics\ (\bibinfo  {publisher} {Springer International Publishing},\
  \bibinfo {year} {2018})\BibitemShut {NoStop}%
\bibitem [{\citenamefont {Andersen}\ \emph {et~al.}(2003)\citenamefont
  {Andersen}, \citenamefont {Bohr}, \citenamefont {Stenum}, \citenamefont
  {Rasmussen},\ and\ \citenamefont {Lautrup}}]{andersen1}%
  \BibitemOpen
  \bibfield  {author} {\bibinfo {author} {\bibfnamefont {A.}~\bibnamefont
  {Andersen}}, \bibinfo {author} {\bibfnamefont {T.}~\bibnamefont {Bohr}},
  \bibinfo {author} {\bibfnamefont {B.}~\bibnamefont {Stenum}}, \bibinfo
  {author} {\bibfnamefont {J.~J.}\ \bibnamefont {Rasmussen}},\ and\ \bibinfo
  {author} {\bibfnamefont {B.}~\bibnamefont {Lautrup}},\ }\href
  {https://doi.org/10.1103/PhysRevLett.91.104502} {\bibfield  {journal}
  {\bibinfo  {journal} {Phys. Rev. Lett.}\ }\textbf {\bibinfo {volume} {91}},\
  \bibinfo {pages} {104502} (\bibinfo {year} {2003})}\BibitemShut {NoStop}%
\bibitem [{\citenamefont {Andersen}\ \emph {et~al.}(2006)\citenamefont
  {Andersen}, \citenamefont {Bohr}, \citenamefont {Stenum}, \citenamefont
  {Rasmussen},\ and\ \citenamefont {Lautrup}}]{andersen2}%
  \BibitemOpen
  \bibfield  {author} {\bibinfo {author} {\bibfnamefont {A.}~\bibnamefont
  {Andersen}}, \bibinfo {author} {\bibfnamefont {T.}~\bibnamefont {Bohr}},
  \bibinfo {author} {\bibfnamefont {B.}~\bibnamefont {Stenum}}, \bibinfo
  {author} {\bibfnamefont {J.~J.}\ \bibnamefont {Rasmussen}},\ and\ \bibinfo
  {author} {\bibfnamefont {B.}~\bibnamefont {Lautrup}},\ }\href@noop {}
  {\bibfield  {journal} {\bibinfo  {journal} {Journal of Fluid Mechanics}\
  }\textbf {\bibinfo {volume} {556}},\ \bibinfo {pages} {121} (\bibinfo {year}
  {2006})}\BibitemShut {NoStop}%
\bibitem [{\citenamefont {Cristofano}\ \emph {et~al.}(2016)\citenamefont
  {Cristofano}, \citenamefont {Nobili}, \citenamefont {Romano},\ and\
  \citenamefont {Caruso}}]{CRISTOFANO}%
  \BibitemOpen
  \bibfield  {author} {\bibinfo {author} {\bibfnamefont {L.}~\bibnamefont
  {Cristofano}}, \bibinfo {author} {\bibfnamefont {M.}~\bibnamefont {Nobili}},
  \bibinfo {author} {\bibfnamefont {G.}~\bibnamefont {Romano}},\ and\ \bibinfo
  {author} {\bibfnamefont {G.}~\bibnamefont {Caruso}},\ }\href
  {https://doi.org/https://doi.org/10.1016/j.expthermflusci.2015.12.005}
  {\bibfield  {journal} {\bibinfo  {journal} {Experimental Thermal and Fluid
  Science}\ }\textbf {\bibinfo {volume} {74}},\ \bibinfo {pages} {130 }
  (\bibinfo {year} {2016})}\BibitemShut {NoStop}%
\end{thebibliography}%


\end{document}